# Leaves Compute

Authors: David Peak [a], Keith A. Mott [b], and Mathew T. Hogan [c]
a Physics Department, Utah State University, Logan, Utah 84322-4415
b Biology Department, Utah State University, Logan, Utah 84322-0300
c Physics and Astronomy Department, University of Utah, Salt Lake City, Utah 84112-0830

Abstract

Control of gas exchange between a leaf's interior and the surrounding air is accomplished by variations in the turgor pressures in the small epidermal and guard cells that cover the leaf's surface. These pressures respond to changes in light intensity and color, temperature, $CO_2$ concentration, and air humidity. The dynamical equations that describe these processes are formally identical to those that define computation in a two-layer, adaptive, cellular nonlinear network. This identification suggests that leaf gas-exchange processes can be understood as a kind of analog computation.

## I. Introduction

"Emergent computation" [1-3], is a type of problem-solving that results from the collective activity of systems of functional units, but in which there is no central unit capable of coordinating the behavior of the others. Leaderless, distributed problem-solving is a central characteristic of biological organisms and also of various biomimetic computational networks. This fact suggests the possibility that there might be situations where biological behavior and computation are inextricably connected. Often—largely because they are able to communicate by various signals with one another—plants have been suggested as examples of such biological-computational systems [4-8]. These proposals have typically not included a precise definition of which computational scenario the plant system is supposed to be part of, though. Here, however, it is argued that the *structure and biological function* of a collection of stomata on the surface of a plant leaf are formally identical to the *structure and computational function* of a two-layer, adaptive, cellular nonlinear network [9].

The argument for this connection is based on a theoretical model that incorporates biologically relevant structures and interactions to explain how stomatal apertures might collectively adjust to external stimuli. The biological relevance of the model is justified in the following by comparing the model's outputs to well-known experimental results. The model is subsequently identified, feature-by-feature, with an actual, hardware, computational device. It is therefore proposed that the manipulation of water resources by (at least some) plants is a specific form of *analog* computation.

## II. Elements of the Stomatal Network Model (the SNM)

The epidermis of a plant leaf is a waxy covering that prevents massive water loss due to evaporation from the leaf's interior. In order for photosynthesis to proceed, the epidermis is perforated with stomata—small, variable aperture openings that allow carbon dioxide to enter the leaf from the surrounding air. When open, stomata also allow water vapor to escape. So, stomatal aperture varies with changing external conditions to acquire sufficient carbon dioxide for photosynthesis while simultaneously preventing desiccation. This feat is accomplished by small "guard" cells that define the stomatal opening pushing against much larger "epidermal" cells in which they are embedded. The mechanical forces required for this process result from turgor pressure differences within the respective cells that are controlled by water transport into and out of the cells. The effect of these events is emergent; they arise from the *collective* action of all of the stomata—there is no leaf tissue that organizes the activity.

Essential for identifying a biological stomatal system with a relevant computational hardware system is the recognition that stomata at different places on a leaf are *not* independent of one another. Experiments in which a single stoma responds to a localized perturbation—such as by blowing dry air on it—show that many surrounding stomata also quickly respond [10]. Additional support for stomatal interaction is the ubiquitous phenomenon of "stomatal patchiness," in which large regions of similar stomatal openness appear to sweep coherently over the surface of a leaf [11]. In other words, stomata are not autonomous entities; they are linked to one another in a kind of network.

(1) The stomatal unit.

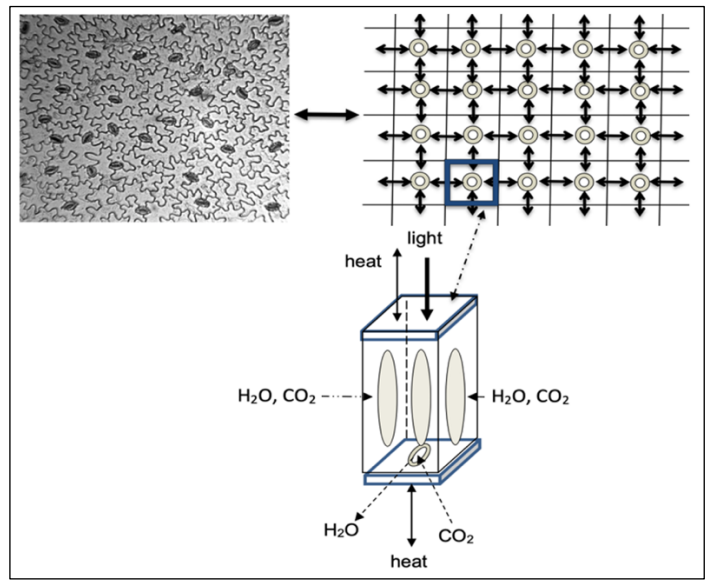

The SNM proposed here assumes the leaf is hypostomatous, i.e., with stomata on the under (abaxial) side only. Often real stomata are roughly regularly distributed. (See the top leftmost image in Figure 1; in it, stomata are "bean-shaped" and epidermal cells appear similar to jigsaw puzzle pieces.) For simplicity, the SNM treats adjacent stomata as evenly spaced and the leaf is envisioned as a regular array of identical, densely packed, rectangular-solid shaped plugs (top rightmost image). In the SNM, each plug has square epidermal caps on top and bottom, with mesophyll tissue filling the interior (lower image).

Figure 1

Each abaxial cap in the model contains one stoma.
This total structure is defined here as a "stomatal unit." The center of each stomatal unit is located at the site $r = (i, j)$ on a square grid, where $i$ and $j$ are integers.

Subsections (2)-(5) which follow discuss experimentally-informed details of the structure and dynamics associated with the SNM. Subsequently, subsection (6) shows that the SNM's dynamic output closely agrees with observed stomatal behaviors.

(2) Energy fluxes

As illustrated in Figure 1, light energy passing through the top (adaxial) cap of each stomatal unit is partially absorbed in the unit's mesophyll. Both caps exchange thermal energy with the surrounding air.

(3) Liquid water

Liquid water flows from the plant's roots (or the leaf's petiole) to the mesophyll cells in the interior of the leaf. Epidermal cells on the abaxial surface of a unit exchange liquid water directly with the surfaces of the unit's mesophyll cells and also with abaxial epidermal cells in adjacent units.

(4) Water vapor and carbon dioxide

Saturated water vapor evaporating from the surfaces of mesophyll cells fills the leaf interior. Stomatal guard cells exchange their internal liquid water with the water vapor in the adjacent "stomatal cavity," a region in the unit's interior close to the stomatal cap. When the stomatal pore is open, carbon dioxide enters the leaf interior and is absorbed in the mesophyll in the process of photosynthesis.

(5) Dynamics

   a. The stomatal aperture

The instantaneous turgor pressure, $P_g(r;t)$, in the guard cells at site $r$ is assumed to be the same in both cells. If that pressure is sufficiently large to overcome the instantaneous turgor pressure, $P_e(r;t)$, in the surrounding epidermal cells, that stoma opens. The instantaneous aperture of the opening, in the SNM, is proportional to the weighted difference, $P_g(r;t) - mP_e(r;t)$, in which the dimensionless factor $m$ is related to the "mechanical advantage" of the larger epidermal cells over the smaller guard cells. The water vapor conductance associated with this aperture is

$$g_w(r;t) = h\{\chi[P_g(r;t) - mP_e(r;t)]\}. \tag{1}$$

In (1), $h$ is a dimensionless *trilinear function* of its argument: that is, (i) the conductance is 0 if the argument is negative; (ii) the conductance equals $g_{w,max}$ if the argument exceeds some maximum value; and (iii) the conductance is just $\chi[P_g(r;t) - mP_e(r;t)]$, otherwise. The positive coefficient $\chi$ is determined by the elasticity of the cells. When $0 < g_w(r;t) < g_{w,max}$,

$$\frac{dg_w(r;t)}{dt} = \chi\left[\frac{dP_g(r;t)}{dt} - m\frac{dP_e(r;t)}{dt}\right]. \tag{2}$$

   b. Summary of the dynamics

The Appendix, section IV below, contains the many details of how, in the SNM proposed here, the cellular turgor pressures in (1) are related to external light intensity, humidity, temperature, and carbon dioxide concentration. In the following, the dynamical model resulting from these details is used to confirm the hypothesis that real stomatal networks might "compute." To accomplish this, the SNM equations are first shown to appropriately describe examples of observed stomatal behavior. Following that, the theoretical stomatal network architecture is shown to be identical to that of a computational network.

After aggregating the various pieces of the SNM developed in the Appendix, the relevant equations for the rates of change of the epidermal and guard cell pressures in (2) become respectively

$$dP_e(r;t)/dt = -\lambda_e P_e(r;t) + \lambda_e\{-\rho E(r;t) + \Gamma_e(r;t)RT(r;t) + \eta_{ee}\sum_{r'}[P_e(r';t) - \Gamma_e(r';t)RT(r';t)]\} \quad (3a)$$

$$dP_g(r;t)/dt = -\lambda_g P_g(r;t) + \lambda_g\left[\Gamma_g(r;t)RT(r;t) + \frac{RT(r;t)}{V_W}\ln\left(\frac{w_c(r;t)}{w_m(r;t)}\right)\right]. \quad (3b)$$

Here, $E(r;t)$ is the evaporative flux from, and $T(r;t)$ is the temperature throughout, the plug at site $r$; $\rho$ is the resistance to liquid water flow from the roots (or petiole) to the leaf; $\lambda_e$ and $\lambda_g$ are rate constants for water transport across the respective cell membranes; $\Gamma_e(r;t)$ and $\Gamma_g(r;t)$ are dissolved ion concentrations in the respective cells; $\eta_{ee}$ measures the strength of water sharing between adjoining epidermal cells; and the $w$ variables in (3b) are water vapor mole fractions in the stomatal cavity and at the mesophyll.

(6) Comparing example output of the SNM with experimental results

In real plants, neighboring stomata typically differ somewhat in orientation and cellular structure (see Fig. 1, for example). To account for this variability in equations (3a) and (3b), the ionic concentrations $\Gamma_e(r;t)$ and $\Gamma_g(r;t)$ are randomly assigned slightly different values from position to position. In the example simulations that follow (resulting by integrating equation (2) after inserting (3a) and (3b)), this small random variability generates small, calculated, aperture fluctuations.

 a. Stomatal patchiness

Often experiments examining stomatal behavior start with a leaf in the dark. When an illumination is turned on, high-resolution thermal imaging initially shows small, randomly positioned, temperature variations over the leaf's surface. After some minutes pass, these small fluctuations often grow into large patches of stomata in which similar warmer temperatures abut other large patches of stomata with similar cooler temperatures. This is the phenomenon of "stomatal patchiness" mentioned previously; it has been identified in many species in both the laboratory and the wild [11]. Patchiness seems to suggest irregular, suboptimal stomatal behavior, so, its ubiquity is puzzling. Nonetheless, the generation of patchiness is a fundamental first test that any plausibly correct model of stomatal arrays must pass. Figure 2 shows that patchiness indeed arises spontaneously from equations (3a), (3b), and (2).

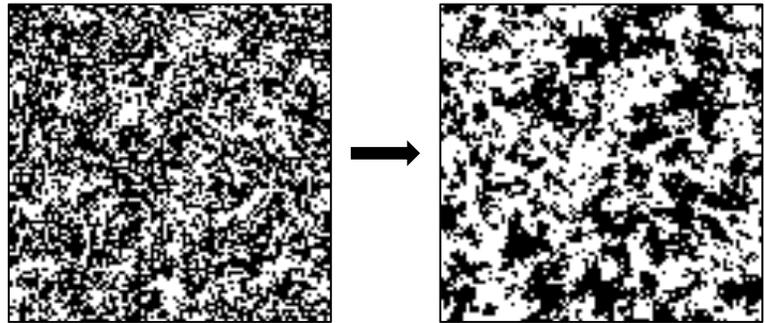

Figure 2

The assumed small random variations in the initial configuration produce, in the left image of Figure 2, small random initial temperature fluctuations—shown as white for "warm" and black for "cool." Several calculated "minutes" later, depicted on the right in Figure 2, the initial fluctuations have organized into much larger patches of warm and cool, similar to what is frequently observed in experiments with real plants.

### b. The wrong-way response

When humidity in the air surrounding an illuminated biological leaf is gradually lowered the leaf's stomata tend to gradually close. If the humidity drop is rapid, however, sometimes the stomata at first briefly open before closing. This is the so-called "wrong-way response," which is observed in many experiments [12]. Figure 3 shows a simulation with the SNM of a rapid humidity drop experiment.

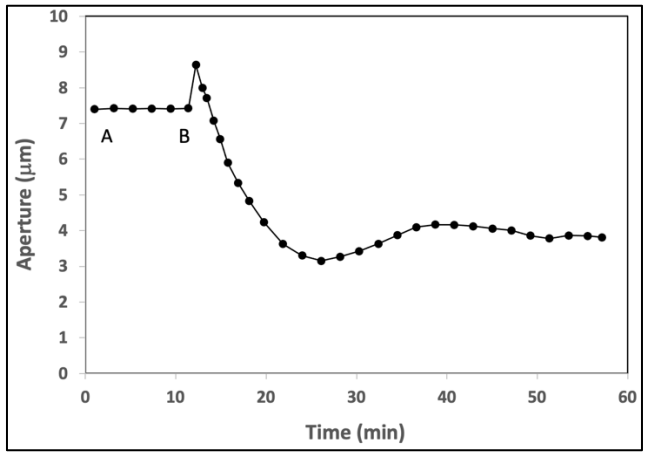

Figure 3

Between times "A" and "B" in the figure, the "leaf" is in steady state with a large, steady, average stomatal aperture. At "B" the simulated external humidity is suddenly reduced. The immediate response is that the stomata *open*—i.e., go the "wrong way"— after which the stomata close, as expected, with eventually a smaller average aperture.

### c. Red and blue light

When real plants, with already open stomata, are exposed to bright monochromatic red light their stomata often begin a long duration episode of large amplitude oscillations. When a small amount of blue light is combined with the red, however, the oscillations are often observed to damp out [13].

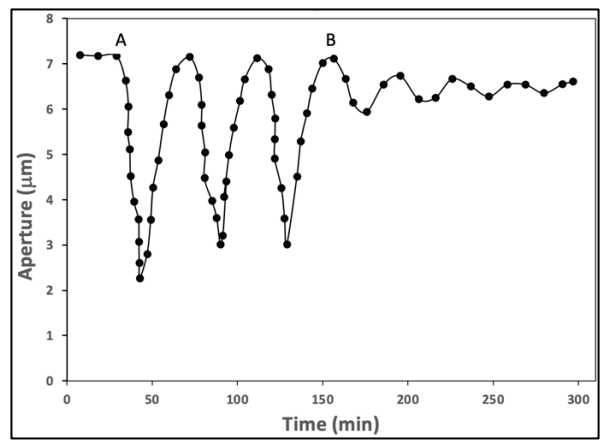

Figure 4

Figure 4 shows that this effect arises naturally in the SNM. At first, the model stomata are steadily open in white light. At "A" the illumination is changed to equal intensity red light, and the model stomata begin oscillating with large amplitude. At "B" 5% of blue light replaces 5% of the red. Subsequently, the oscillations diminish.

d. Stomatal oscillations in the dark

Under some conditions, the stomata of at least some plants can be slightly open in the dark. When that occurs, typically the stomata respond to even very small environmental changes by rapidly and persistently oscillating [14]. If, in the SNM, the stomatal pores are slightly open in the dark, the model also exhibits small, persistent, rapid, and irregular stomatal oscillations. This is shown in the SNM simulation in Figure 5.

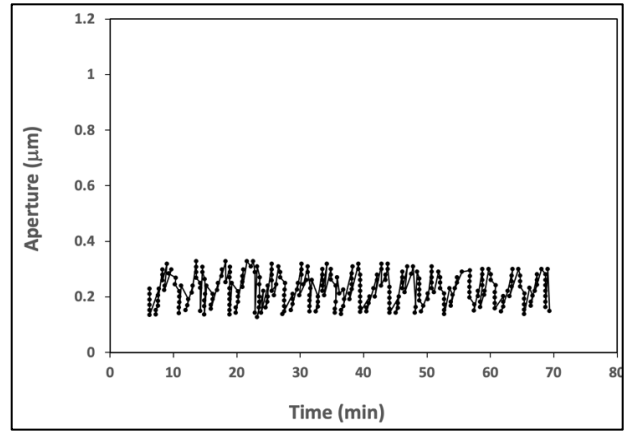

Figure 5

e. Water use efficiency

Figure 6 depicts the time course of the SNM simulation of water use efficiency [15]. WUE on the graph is defined as the ratio of the rate at which carbon dioxide is acquired to the rate at which water is lost through open stomata. At first, the model stomata are closed in the dark. After light is turned on stomata gradually open (as in Figure 2) and WUE increases. Eventually, WUE reaches an approximate asymptotic value. This value is a maximum for the specific plant and the starting environmental conditions.

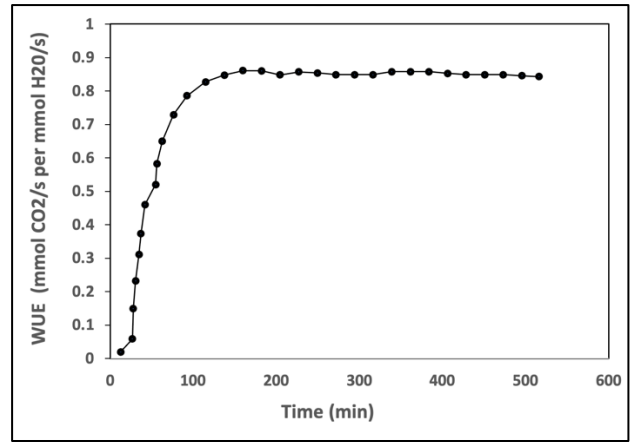

Figure 6

f. The simulations are not tailored to fit the experiments

The example SNM results shown in Figures 2-6 are not individually "put in" with different parameters for each different example. All arise after iteratively adjusting the model parameters so that the various outputs for the different external conditions—*using a single set of parameter values*—mimic typical experimental observations found in many species.

g. The SNM yields a new view of the ubiquity of stomatal patchiness

In the SNM presented above, increased water use efficiency and the growth of stomatal patchiness are strongly correlated. In Figure 6 water use efficiency is very low shortly after a leaf is exposed to bright light. In that condition, the stomatal configuration on the leaf is similar to what is depicted in the left image in Figure 2—i.e., a random scrambling with no large patches. As patchiness grows—for example as in the right image in Figure 2—so too does water use efficiency, as shown for later times in Figure 6.

This correlation occurs in the SNM only when adjacent epidermal cells easily exchange water with one another, that is, when the model parameter $\eta_{ee} \approx 1/4$ (i.e., strong epidermal cell water exchange). If, on the other hand, $\eta_{ee} \approx 0$ (weak water exchange), no large patches form. In the latter case, the small scale irregularity in the initial open-closed configuration persists and so does low water use efficiency.

Patchiness, and therefore increased water use efficiency, appear to be natural consequences of the evolution, for different plants, of increased water exchange between irregularly shaped epidermal cells. Thus, the seemingly "puzzling" ubiquity of stomatal patchiness might actually be what should be expected.

III. Two-layer, adaptive, cellular nonlinear networks—and implications for stomata

A two-layer, adaptive, cellular nonlinear network paradigm was proposed in 2007 by Koeppl and Chua [9]. Their model is a computational network of electronic "cells," each of which is capable of changing its internal voltage depending on its instantaneous value as well as the voltages of cells to which it is wired. Each cell broadcasts its internal state to its *nearest neighbors* only. In the adaptive version, each layer of cells can receive signals from external sources. Depending on the program being executed, each cell can change its internal state according to its current state and all of its input states, and subsequently broadcast the result to the cells to which it is connected. Eventually, this activity settles into a steady (or steadily oscillating) state, which corresponds to "an answer" to the problem being solved. In this version, the problem to be solved can change when the external inputs change; that is, the network can "adapt."

The dynamical equations in the Koeppl and Chua (K-C) adaptive CNN are summarized by

$$\frac{dx^l(r;t)}{dt} = -\lambda_l x^l(r;t) + A^l[u(r';t); r] \cdot y^l(r';t)$$

$$+ \sum_{r'} B^l[u(r';t); r] \cdot u^l(r';t) + z^l(r;t). \qquad (4)$$

Here: $x^l(r;t)$ is the internal state of the cell at position $r = i,j$ and time $t$ in layer $l = 1$ or 2; $y^l(r;t)$ is the output state of the cell at position $r = i,j$ and time $t$ in layer $l$; $y^l(r;t) = F[x^l(r;t)]$, where $F$ is a *trilinear function* of its argument; $u^l(r;t)$ is a function of the external inputs to the cell at position $r$ and time $t$ in layer $l$; the sums are over all the nearest neighbor positions of $r$, as well as $r$ itself; $\lambda_l$ is a "relaxation" rate constant for cells in layer $l$; the $A$s and $B$s are functions of the external inputs (thus, allowing for adaptability); $z^l(r;t)$ is the "bias" of the cell at $r$ and time $t$ in layer $l$.

The dynamical equations, (3a) and (3b) for the SNM can be understood as such an adaptive CNN, with the indices $e$ and $g$ corresponding to the leaf's computational "layers." Identifying the various SNM terms in (3a) and (3b) with those in the K-C multilayer adaptive CNN (4) yields:

For the $e$–computational layer:
- $P_e \leftrightarrow$ the CNN state $x^1$;
- $\lambda_e(1 + 4 \cdot \eta_{ee}) \leftrightarrow$ the CNN relaxation rate constant $\lambda_1$;
- $\lambda_e\{-\rho g(r;t)[w(r;t) - w_a]\} \leftrightarrow$ the CNN $A^1$ term;
- $\lambda_e[\Gamma_e(r;t)RT_e(r;t) - \eta_{ee}\sum_{r'}\Gamma_e(r';t)RT_e(r';t)] \leftrightarrow$ the CNN $B^1$ term;
- $\lambda_e\eta_{ee}\sum_{r'}P_e(r';t) \leftrightarrow$ the CNN $z^1$ term.

For the $g$–computational layer:
- $P_g \leftrightarrow$ the CNN state $x^2$;
- $\lambda_g \leftrightarrow$ the CNN relaxation rate constant $\lambda_2$;
- $\lambda_g\left[\Gamma_g(r;t)RT_e(r;t) + \frac{RT_e(r)}{v_W}\ln(\frac{w_c(r;t)}{w_{sat}(r;t)})\right] \leftrightarrow$ the CNN $B^2$ term;
- the CNN $A^2$ and $z^2$ terms are zero.

Conclusion: **The stomatal dynamical system is structurally identical to a 2–layer, adaptive, cellular nonlinear network which computes.**

Note that when it is suitably parametrized, a CNN can simulate a corresponding leaf's stomatal network behavior. Therefore, a CNN can be used to perform large numbers of "what-if" experiments in a short time, investigating "yields and properties" of possibly hitherto unknown "plants." **Determining which CNN parameters lead to improved numerical outputs can then be used to suggest possible genetic modification and/or environmental control techniques aimed at improving crop yields, water use, and other potentially desirable characteristics of plants in the real world**.

IV. Appendix: The various details underlying Equations (3a) and (3b)

The equations developed below tacitly assume that changes in external factors are sufficiently small that possible chemical effects due to plant hormones, for example, can be ignored.

(1) Rates of change of turgor pressure

Cellular turgor pressure varies with a cell's water content. When free to do so, water flows from higher *water potential* to lower. Thus, the turgor pressure in the guard cells at site $r$ varies in time as

$$\frac{dP_g(r;t)}{dt} = \lambda_g[\Psi_c(r;t) - \Psi_g(r;t)], \tag{A1}$$

where $\Psi_g$ is the water potential of the liquid inside the guard cells and $\Psi_c$ is the water potential of the vapor in the stomatal cavity adjacent to the guard cells; $\lambda_g$ is the rate constant for water transport in and out of guard cells. (In the SNM presented here, guard cells only exchange water with the vapor in their adjacent stomatal cavity.)

Similarly, the turgor pressure in the epidermal cells on the abaxial surface of the leaf (i.e., where the stomata are) at site $r$ varies in time as

$$\frac{dP_e(r;t)}{dt} = \lambda_e[\{\Psi_m(r;t) + \eta_{ee}\sum_{r'}\Psi_e(r';t)\} - \Psi_e(r;t)]. \tag{A2}$$

In this expression, $\Psi_m$ is the water potential of the liquid on the mesophyll cells; $\Psi_e$ is the water potential of the liquid inside epidermal cells; $r'$ is the site of a nearest neighbor of $r$; and $\eta_{ee}$ is the probability that abaxial epidermal neighbors share water.

(2) Water potentials

If the liquid water potential at the *source* (the petiole for a detached leaf, the roots for a potted plant) is defined as $\Psi_S = 0$, then the liquid water potential at the mesophyll is

$$\Psi_m(r,t) = -\rho E(r,t), \tag{A3}$$

where $E(r,t)$ is the transpiration rate at time $t$ from the stoma at $r$, and $\rho$ is the resistance to liquid water flow from the source to the leaf interior.

The liquid phase water potentials within the epidermal and guard cells are the respective differences between turgor ($P$) and osmotic ($\Pi$) pressures. The osmotic pressures are of the form $\Pi_x(r;t) = \Gamma_x(r;t)RT(r;t)$, where $\Gamma_x(r;t)$ is the relevant (i.e., $x = e$ or $g$) ionic concentration at $r;t$; $R$ is the universal gas constant; and $T(r;t)$ is the leaf temperature in Kelvins at $r;t$. Thus, for both epidermal and guard cells,

$$\Psi_x(r;t) = P_x(r;t) - \Gamma_x(r;t)RT(r;t). \tag{A4}$$

In general, the water potential of a vapor of arbitrary water content in air—with mole fraction $w(r;t)$ (mmol-water/mol-air)—in equilibrium with liquid water at temperature $T$ is given by

$$\Psi[T(r;t)] = \frac{RT}{V_W}\ln\left[\frac{w(r;t)}{w_m[T(r;t)]}\right], \tag{A5}$$

where $V_W$ is the molar volume of pure water and $w_m[T(r,t)]$ is the saturated water vapor mole fraction over liquid water at temperature $T(r;t)$.

(3) Ionic concentrations

In the SNM, the ionic concentration in solution within all *epidermal* cells is taken to be constant: $\Gamma_e(r;t) = \Gamma_{e0}$. On the other hand, the ionic concentration in solution within *guard cells* is partly fixed by metabolism, $\Gamma_{g0}$, but also varies with the color and intensity of the incident light. In particular, guard cells have an intrinsic response to absorbed blue light and, in addition, receive—and respond to—a light-related "signal" generated by light absorbed in the mesophyll. Thus, in the SNM, the ionic concentration in solution within the guard cells is

$$\Gamma_g(r;t) = \Gamma_{g0} + \Gamma_b(r;t) + \Gamma_s(r;t), \tag{A6}$$

where $\Gamma_b(r;t) = \Gamma_{b0} \frac{I_b}{I_b+K_b}$ and $\Gamma_s(r;t) = \Gamma_{s0} \frac{I_0}{I_0+K_s c_i(r;t)}$.

In (A6), $\Gamma_{g0}$ is independent of site; $I_b$ is the blue light intensity and $I_0$ is the total light intensity (both assumed to be spatially uniform in experiments) at site $r$ and time $t$, and the $K$'s are appropriate rate constants. The quantity $c_i(r;t)$ is the instantaneous mole fraction of gaseous carbon dioxide within the leaf at $r;t$. The SNM assumes that the amount of "signal" (possibly a positive ion [16]) delivered to the guard cells is proportional to the light energy *not* used in photosynthesis. For a given intensity of light, therefore, the less (alternatively, more) $CO_2$ available, the more (less) signal is produced. This causes the water potential in the guard cells to decrease (increase) and therefore water to enter (exit), and, as a result, the stomatal aperture to increase (decrease).

(4) Temperature

Experiments equipped with high resolution thermal imaging capability often observe large patches of different temperatures moving *coherently* on a leaf's surface over time scales of minutes. Sometimes such patches can persist for hours. This suggests that lateral thermal conductivity on the surface of and within a leaf is small. Therefore, the energy balance at site $r$ involving the leaf, the surrounding air, and the incident light leads to a unit-average *local leaf temperature* of the form

$$T(r;t) = T_a + \frac{\delta \cdot I_0 - L_W E(r;t)}{K_a}, \qquad (A7)$$

where $T_a$ is the air temperature, $\delta$ is the fraction of the incident light intensity, $I_0$, absorbed by the mesophyll that is dissipated as heat, $L_W E(r;t)$ is the rate of cooling of a unit due to transpiration ($L_W$ is the latent heat of vaporization), and $K_a$ is the thermal conductivity of air. (Thermal radiation at typical air temperatures is assumed to be negligible.)

(5) Water vapor

The water vapor *inside* the leaf adjacent to the mesophyll is assumed to be saturated at the temperature given by equation (A7) above. The water potential at the mesophyll is given by equation (A3) above, but, for most laboratory conditions (where source-to-leaf distances are small), at least, is well-approximated as 0. In that case, the mole fraction of the saturated vapor at the mesophyll is the usual saturated vapor pressure at the temperature $T(r;t)$: $w_m(r;t) = w_{sat}[T(r;t)]$.

If the stomatal pore is open, water vapor diffusion from the mesophyll to the surrounding air requires that the water vapor mole fraction in the stomatal cavity be less than $w_m(r;t)$. A reasonable assumption is

$$w_c(r;t) = \sigma w_a + (1-\sigma)w_m(r;t), \qquad (A8)$$

where $w_a$ is the water vapor mole fraction in the surrounding air and $\sigma$ is a small fraction [17] (zero, if the pore is closed). In this approximation, the transpiration rate from site $r$ is

$$E(r;t) = g_W(r;t)(1-\sigma)[w_m(r;t) - w_a]. \qquad (A9)$$

Though small, $\sigma$ determines how the water potential in the stomatal cavity varies with the humidity in the air, and this in turn affects the epidermal turgor pressure (through equation (A1))—and therefore the conductance, $g_W(r;t)$.

(6) Carbon dioxide

As discussed with respect to equation (A5) above, the ionic concentration in the liquid solution in the guard cells varies with the concentration of carbon dioxide in the air inside the leaf. To determine the latter requires a kinetics equation describing the leaf processes involving $CO_2$.

It is well known [18] that stomatal movements are much slower than photosynthetic-related processes. Thus, the $CO_2$ kinetics equation can be taken to be in steady state compared with pressure changes of guard and epidermal cells due to water exchange. Consequently,

$$0 = g_c(r;t)[c_a - c_i(r;t)] - k_c c_i(r;t) I_0 + \lambda_c [\tfrac{1}{4}\sum_{r'} c_i(r';t) - c_i(r;t)]. \quad (A10)$$

The terms in (A10) describe how the carbon dioxide in the interior of a stomatal unit participates in different interactions. The first term on the RHS represents the diffusive uptake of $CO_2$ from the air; the second term approximates how gaseous $CO_2$ is incorporated into photoproducts; and the third term relates to the possible exchange of gaseous $CO_2$ between neighboring stomatal units. Equation (A10) can be rearranged to yield

$$c_i(r;t) = \frac{g_c(r;t) c_a + \lambda_c \tfrac{1}{4}\sum_{r'} c_i(r';t)}{g_c(r;t) + k_c I_0 + \lambda_c}. \quad (A11)$$

In (A10) and (A11), $g_c(r;t) = 0.6 g_w(r;t)$, due to the larger mass of $CO_2$ relative to $H_2O$.

All of the phenomena described in the subsections of this addendum are incorporated in equations (3a) and (3b).

The Python code and numerical values of the associated parameters used in the simulations shown in Figures 2-6 can be found at https://github.com/MHogan17/LeafModel.